\documentclass[runningheads]{llncs}

\usepackage[spanish]{babel}

\usepackage{graphicx}
\usepackage{float}
\usepackage{tikz}
\usetikzlibrary{shapes.geometric, arrows}
\usepackage{multirow}

\begin{document}
\title{The Importance of Open Data Policy to Tackle Pandemic in Latin America}
\titlerunning{Open Data, Pandemic, Latin America}
\author{Josimar E. Chire Saire\inst{1}}
\authorrunning{Chire Saire, J}
%
\institute{Institute of Mathematics and Computer Science (ICMC),\\  University of São Paulo (USP), Sao Carlos, SP, Brazil\\ 
}

\maketitle              
\begin{abstract}

Open Data Policies can provide transparency, impulse innovation and citizenship participation. Access to the right data in right time can produce huge benefits to population. But, in Latin America there is not enough interest from governments to promote and use properly. By the other hand, global pandemic has caused many damages in different levels, i.e. Economy, Public Health, Education, etc. The paper opens a discussion about the importance of Open Data Policy to mitigate the impact of Covid-19 and overpass this problem.

\end{abstract}

\section{Introduction}

Governments constantly collect data through different institutions or agencies about Health, Economy, Social and other areas. During the last decades, the access to this data was restricted to specific entities. However, governments noticed the open access to this data can produce benefits \cite{opendatagov}: transparency, promote innovation, participation of the citizenship.

There are initiatives to purpose a metric to analyze the previous concept explained, i.e. Open Data Index (GODI). GODI \cite{opendataint} was purposed by Open Knowledge Network, to evaluate openness: government budget, national statistics, national laws, elections, living conditions (air, water, weather), maps and others. It is important to notice that Public Health is not included.

Policymakers can face many challenges when they make a proposal, for many reasons: complexity of problem, information access, constraints, etc. There are many methodologies to make a proposal \cite{Collins2005}, the main concern in health policies are the outcomes or effects on the population. For example, the steps purposed by Dunn \cite{Dunn1981}: definition, prediction,   prescription, description and evaluation.

Global pandemic originated by Covid-19 started at the end of year 2019, declared a pandemic and a global threat by the World Health Organization (WHO) on 11-03-2020 \cite{whowebpage}. WHO\cite{whoplan} defined some measures in the absence of an approved cure, to control the spread of the virus. Many governments\cite{Chire2020} soon adopted and implemented these measures in their own countries, they include physical distancing between persons, wearing face masks, avoiding physical interactions like handshaking, and hugging, and lockdowns, etc.

Latina America \cite{Chire20201} had the first reported cases in Brazil and Mexico, on 26 and 27 February 2020 respectively. A continent consisting of mostly developing countries, Latin America faces  a  unique  challenge  in  dealing  with  the  pandemic. A lack of strong Public Health sector, produced limited capability of testing, slow and inefficient reaction to control the spread of the virus. The primary measure was to set a partial or fully lockdown, closing schools, universities to open online teaching. A lack of good administration of approved budget to improve infrastructure of Public Health produced a not well prepared Health Care System to deal with a increasing number of cases. Finally, it is important to highlight every country take it's own decisions and set policies to combat Covid-19 consequences.

This paper wants to open the discussion about the importance of Open Data Policies in Latin America. The strong relationship between Open Data and Public Health, by consequence, the benefits of having access to relevant to create appropriate policies to benefit the population and in this specific context, reduce the impact of Covid-19 and tackle the pandemic

\section{Open Data Availability}

This sections discusses the Open Data Availability considering the existence of national Statistics, Open Data Policy foundation and the global results of GODI.

\subsubsection{National Statistics}

GODI indicator considers, statistics about demography, economy, i.e. Gross Domestic Product, unemployment and population statistics. The next features were evaluated:
Open License, open and machine readable format, download at once, up to date, publicly available, free of charge. After a search to find National Statistics Institute for each country in South America, the table Tab. \ref{tab:sa} is presented. Names are similar, but the year of foundation is different, in some countries the creation time is less than 50 years. And this data is not easy to find, Venezuela and Paraguay do not have this data in their websites.

\begin{table}[]
\label{tab:sa}
\resizebox{0.75\textwidth}{!}{
\begin{minipage}{\textwidth}
\centering{

\begin{tabular}{llll}
\textbf{Country} & \textbf{Institute/Organization}                                                                                 & \textbf{Foundation} & \textbf{Link}                      \\
Argentina        & \begin{tabular}[c]{@{}l@{}}Instituto Nacional de Estadística y Censos \\ de la República Argentina\end{tabular} & 1968 \cite{far}               & https://www.indec.gob.ar/          \\
Bolivia          & Instituto Nacional de Estadística                                                                               & 1976 \cite{fbo}               & https://www.ine.gob.bo/            \\
Brazil           & Instituto Brasileiro de Geografia e Estatística                                                                 & 1934 \cite{fbr}                & https://www.ibge.gov.br/           \\
Chile            & Instituto Nacional de Estadísticas                                                                              & 1843 \cite{fcl}                & https://www.ine.cl/                \\
Colombia         & Departamento Administrativo Nacional de Estadística                                                             & 1953 \cite{fco}                & https://www.dane.gov.co/           \\
Ecuador          & Instituto Nacional de Estadística y Censos                                                                      & 1976 \cite{fec}                & https://www.ecuadorencifras.gob.ec \\
Paraguay         & Instituto Nacional de Estadística                                                                               & $\sim$1943 \cite{fpy}         & https://www.ine.gov.py/            \\
Peru             & Instituto Nacional de Estadística e Informática                                                                 & 1975 \cite{fpe}                & https://www.inei.gob.pe/           \\
Uruguay          & Instituto Nacional de Estadística                                                                               & 1994 \cite{fuy}                & https://www.ine.gub.uy             \\
Venezuela        & Instituto Nacional de Estadística                                                                               & NA                   & http://www.ine.gov.ve/            
\end{tabular}}
\end{minipage}}
\end{table}

Then, considering the previous table is possible to affirm that some countries noticed the relevance of having a Statistics Center, the benefits and need. Bolivia, Ecuador, Uruguay are the newest, this could imply maturing process to consolidate organization and objectives. Brazil, Chile, Colombia are the oldest, this should mean the organization could be more mature.

\subsection{Open Data Policy}

A second point to consider: the existence of open data website to share data and when this started per country. Table \ref{tab:saod} presents the year of starting the Open Data Policy and the available link to access it.

\begin{table}[hbpt]
\label{tab:saod}
\caption{Open Data Foundation}
\centering{
\begin{tabular}{lll}
\textbf{Country} & \textbf{Foundation} & \textbf{Link}                     \\
Argentina        & 2016                & http://datos.gob.ar/              \\
Bolivia          & NA                  & https://datos.gob.bo/             \\
Brazil           & 2011                & https://dados.gov.br/             \\
Chile            & 2008                & https://datos.gob.cl/             \\
Colombia         & 2014                & https://www.datos.gov.co/         \\
Ecuador          & 2020                & https://sni.gob.ec/datosabiertos  \\
Paraguay         & 2014                & https://www.datos.gov.py/         \\
Peru             & 2017                & https://www.datosabiertos.gob.pe/ \\
Uruguay          & 2015                & https://www.gub.uy/datos-abiertos \\
Venezuela        & NA                  & NA                               
\end{tabular}}
\end{table}

Considering the previous table, is possible to notice countries like Chile, Brazil, Colombia and Paraguay are the oldest and the newest are: Peru, Argentina, Uruguay. All this data were results of searching using: "politica datos abiertos"(open data policy) + country, and there was not data available for Venezuela.

\subsection{GODI Results}

\begin{figure}[H]
\centerline{
\includegraphics[width=0.7\textwidth]{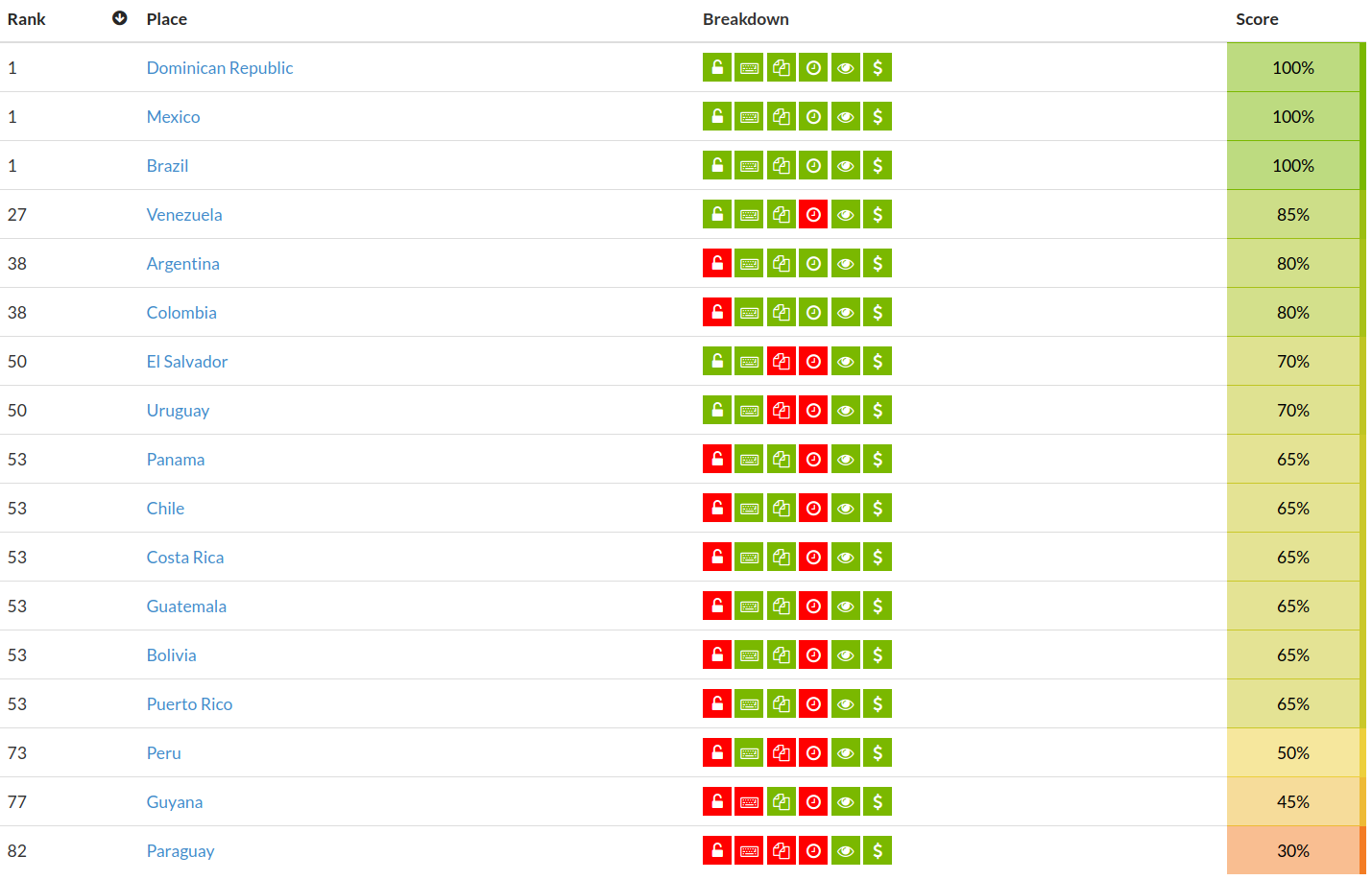}
}
\caption{GODI: National Statistics \protect\footnotemark}
\label{fig:godi}
\end{figure}

In this subsection are presented the results of GODI index. First, the figure Fig. \ref{fig:godi} presents the results of criterion: National Statistics, focusing in Latin America countries. It is possible to notice, the top 5 countries are: Dominican Republic, Mexico, Brazil, Venezuela, Argentina. And the last 5 countries: Bolivia, Puerto Rico, Peru, Guyana, Paraguay. 

\footnotetext{The graphic was created using the available table, Source: https://index.okfn.org/dataset/statistics/}

It is important to highlight that six criterion were analyzed: Open licensed, in an open and machine-readable format, downloadable at once, up to date, publicly available, available free of charge.

Next figure \ref{fig:latotal} presents the available countries considered in the study. The top 5 are: Brazil, Mexico, Colombia, Argentina and Uruguay. At the bottom are: Dominican Republic, Panama, Costa Rica, Guyana, Venezuela.

\begin{figure}[H]
\centerline{
\includegraphics[width=0.7\textwidth]{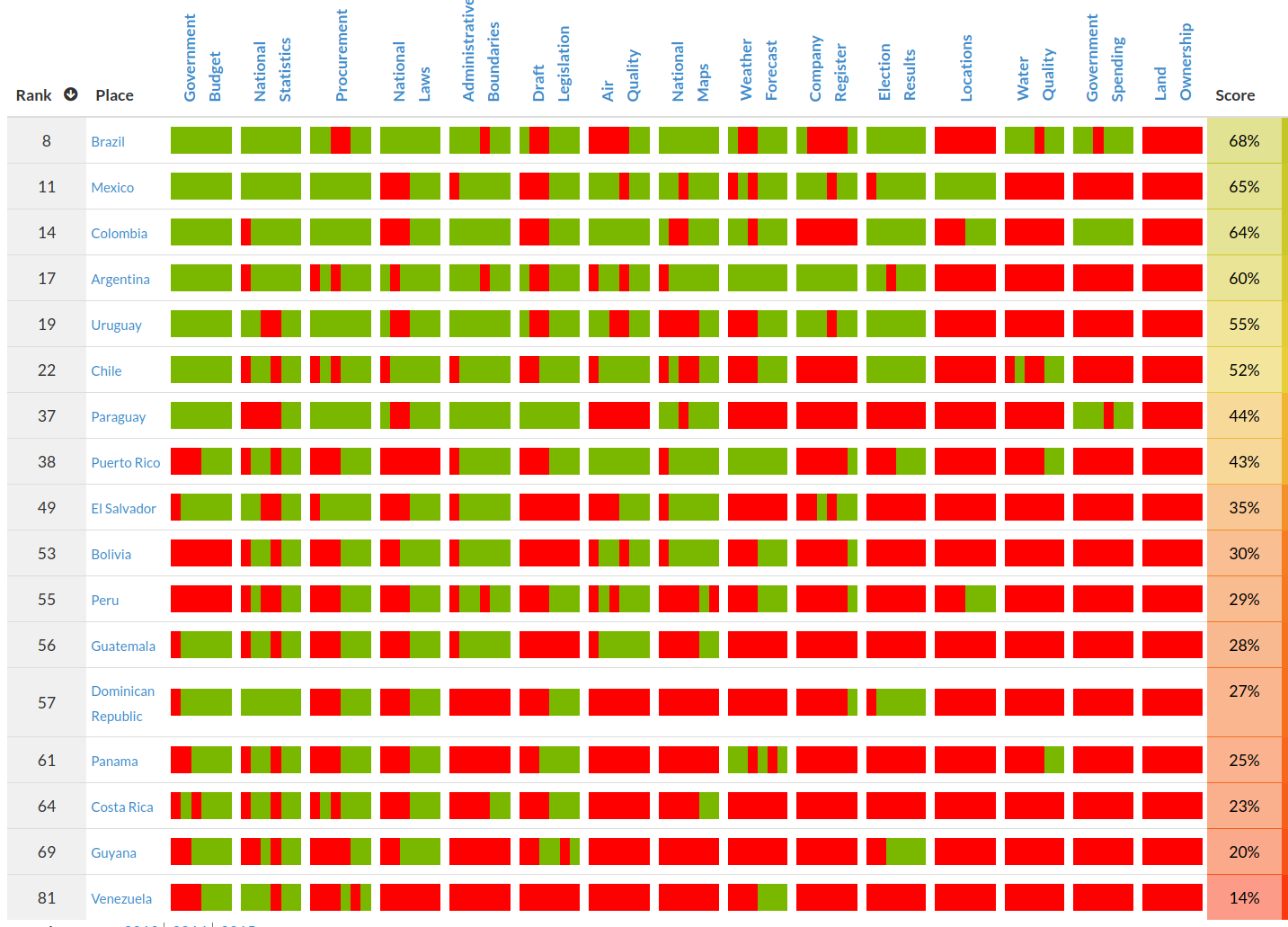}
}
\caption{GODI Indicators \protect\footnotemark}
\label{fig:latotal}
\end{figure}
\footnotetext{The graphic was created using the available table, Source: https://index.okfn.org/place/}

\section{Covid-19 Statistics}

The next figure Fig. \ref{fig:wmtla} presents the statistics about Covid-19 of Latin American countries. The top 5 countries with more cases: Brazil, Argentina, Colombia, Mexico, Peru and the last 5 are: El Salvador, French Guyana, Suriname, Guyana, Nicaragua.

\begin{figure}[H]
\centerline{
\includegraphics[width=0.8\textwidth]{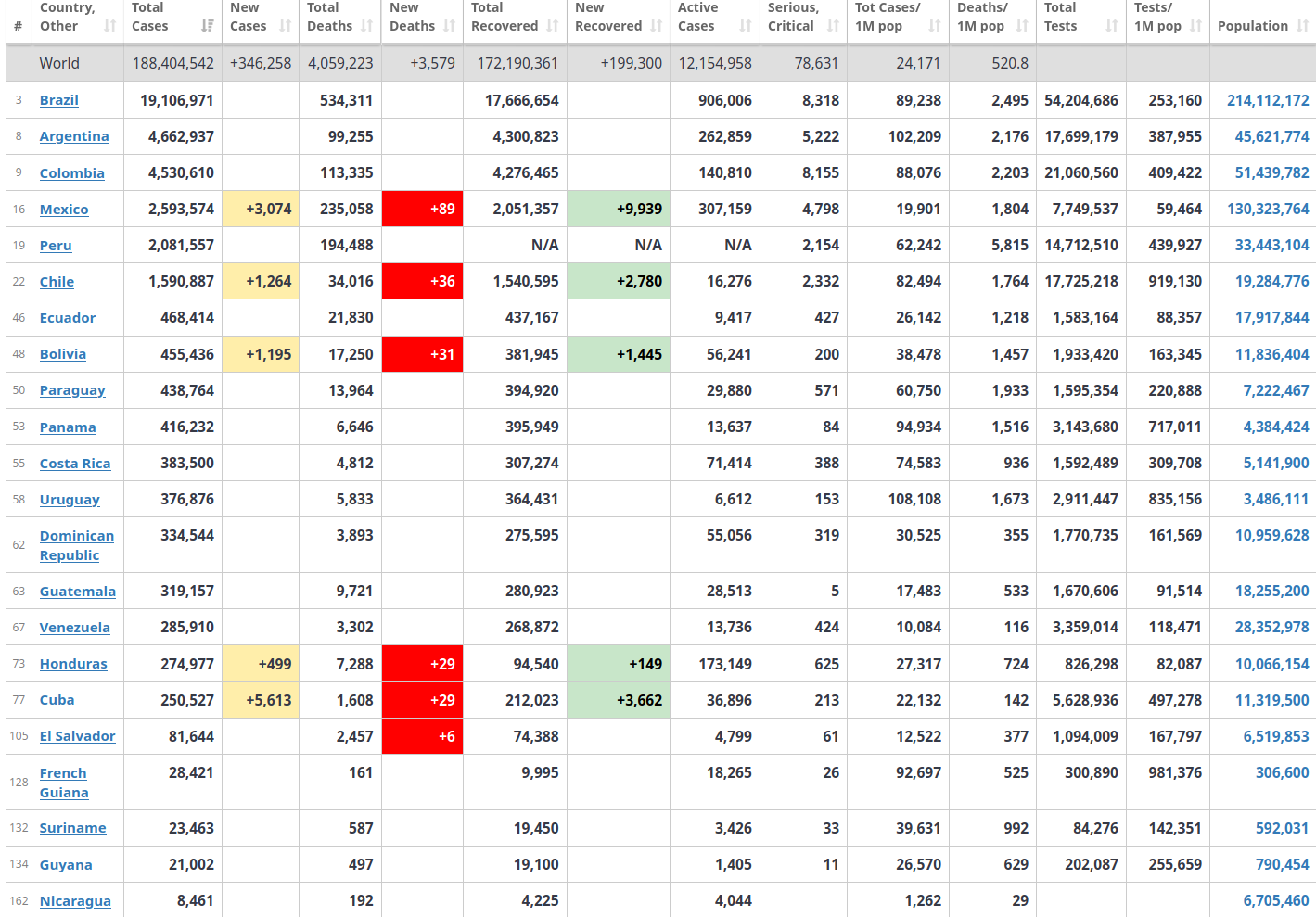}
}
\caption{Covid-19 Statistics of Latin American Countries}
\label{fig:wmtla}
\end{figure}

Next figure Fig. \ref{fig:tc} presents the data considering total cases per country.

\begin{figure}[H]
\centerline{
\includegraphics[width=0.8\textwidth]{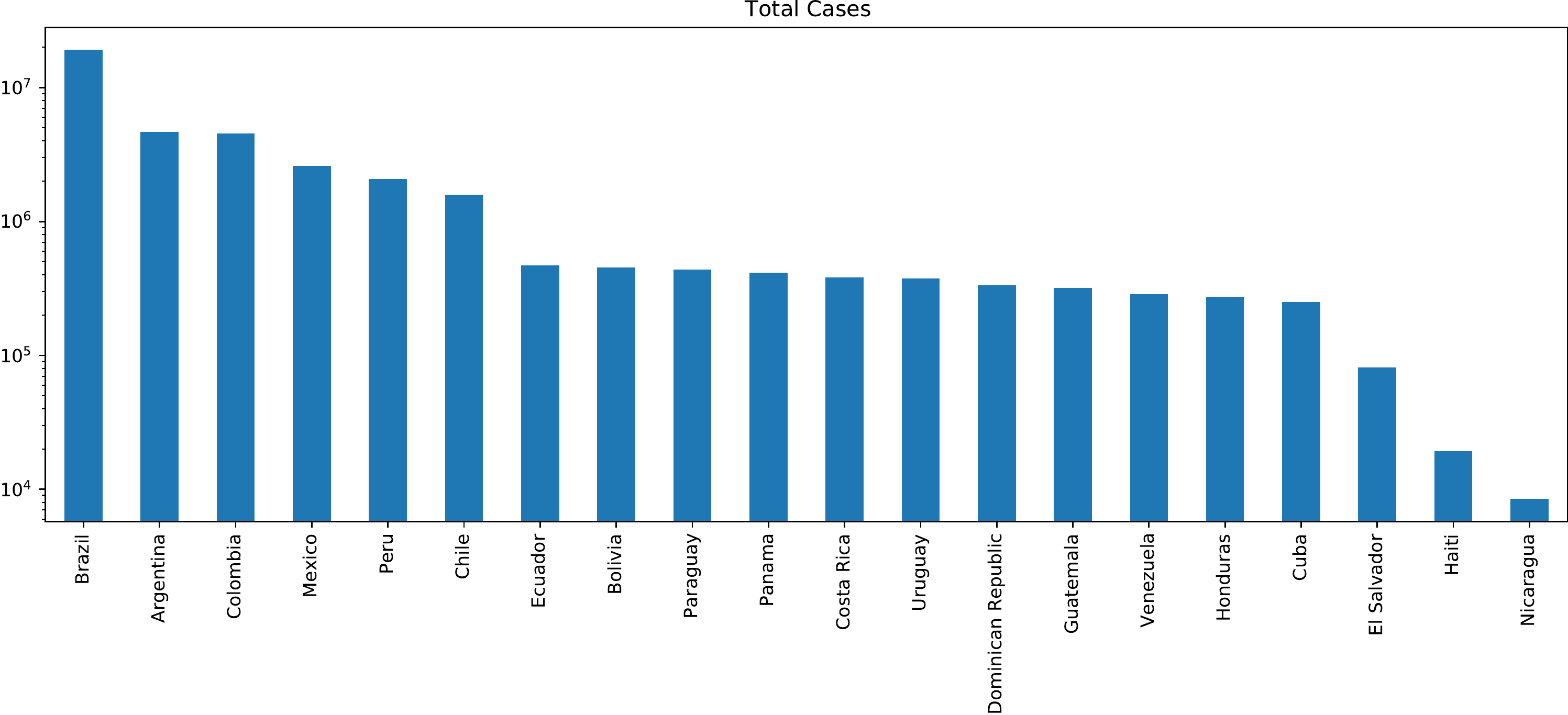}
}
\caption{Total Cases}
\label{fig:tc}
\end{figure}

Again the next countries are in the top: Brazil, Argentina, Colombia, Mexico, Peru. 

\subsection{Discussion}

The discussion starts with the next question: the existence of Open Data Policies can mitigate, help to combat Covid-19 pandemic? The discussion starts with South America focus, but this can be extended for Latin American.
The start point is if there is information at the right time, then policymakers can work properly and reduce/mitigate the impact of Covid-19 in each country in Latin America. Summarizing all the tables and data presented previously, the next figure Fig. \ref{fig:comp} presents the National Statistics foundation year, Open Data Policy foundation and ratio (total cases/population).

\begin{figure}[H]
\centerline{
\includegraphics[width=0.8\textwidth]{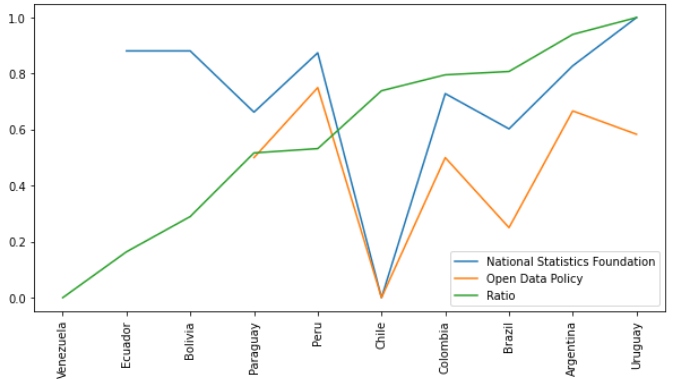}
}
\caption{Comparison}
\label{fig:comp}
\end{figure}

In spite of an oldest foundation in Chile, the ratio is a medium level around 8.24\% population got infected by Covid-19 and Brazil, got around 8.92\%. And, Colombia has 8.8\% ratio. Finally, considering figure \ref{fig:latotal}, Brazil, Colombia and Chile are in the top 10. Then, Open Data Policies is enough to face this kind of challenges as a pandemic, this a question. In spite of, the previous results is important to mention the importance of Open Data Policy to support work of policymakers, and other question is open: what other factors are important to combat efficiently this pandemic? We have not explored the citizenship, industry participation yet. And, the existence of National Statistics and Open Data Policy guaranteed the proper collect of data about social, economical conditions of the population. Besides, in the specific situation of this pandemic the access to data can open new opportunities to create, develop tools can monitor, track population, have a global sight of the situation to have access to vaccine, i.e. Covax and get help from other countries.

\section{Conclusions}

\begin{itemize}
    \item Most of the countries of Latin American have National Statistics, but this paper does not explore the frequency of the data collection and other factors that can make useful this data.
    \item Many countries in South America have create a Open Data policy, Venezuela is an exception, because the url is not available.
    \item Countries with oldest foundation data of National Statistics and Open Data policy got high rates of Covid-19, i.e. Chile, Brazil.
    \item The public availability of data about one country can contribute to create/produce innovation and facilitate vaccine initiatives as Covax.
    \item Other tools can be used to get real time data about population, i.e. Infodemilogy based in Social Networks analysis.
\end{itemize}

%
%

\bibliographystyle{splncs04}
\bibliography{biblio.bib}

\begin{thebibliography}{10}
\providecommand{\url}[1]{\texttt{#1}}
\providecommand{\urlprefix}{URL }
\providecommand{\doi}[1]{https://doi.org/#1}

\bibitem{Chire20201}
Chire-Saire, J., Mahmood, K.: Hope amid of a pandemic: Is psychological
  distress alleviating in south america while coronavirus is still on surge?
  (2020)

\bibitem{Collins2005}
Collins, T.: Health policy analysis: a simple tool for policy makers. Public
  Health  \textbf{119}(3),  192--196 (2005).
  \doi{https://doi.org/10.1016/j.puhe.2004.03.006},
  \url{https://www.sciencedirect.com/science/article/pii/S003335060400099X}

\bibitem{fco}
Dane: Generalidades. Accessed on 11 July 2021 in
  URL:https://www.dane.gov.co/index.php/acerca-del-dane/informacion-institucional/generalidades

\bibitem{opendatagov}
Foundation, O.K.: Welcome to open government data. Accessed on 11 July 2021 in
  URL:{http://opengovernmentdata.org/}

\bibitem{fbr}
Ibge: O ibge. Accessed on 11 July 2021 in
  URL:https://www.ibge.gov.br/acesso-informacao/institucional/o-ibge.html

\bibitem{far}
Indec: Ley 17622. Accessed on 11 July 2021 in URL:
  https://www.indec.gob.ar/tp/documentos/Ley\_17622.htm

\bibitem{fpy}
Ine: Accessed on 11 July 2021 in URL:https://www.ine.gov.py

\bibitem{fbo}
Ine: Dl 14100 ley del sistema nacional de información estadística. Accessed
  on 11 July 2021 in
  URL:https://www.ine.gob.bo/index.php/wpfd\_file/dl-14100-ley-del-sistema-nacional-de-informacion-estadistica/

\bibitem{fcl}
Ine: Historia. Accessed on 11 July 2021 in
  URL:https://www.ine.cl/institucional/nuestra-institucion/historia

\bibitem{fuy}
Ine: Marco jurídico. Accessed on 11 July 2021 in
  URL:https://www.ine.gub.uy/web/guest/ley\-16616

\bibitem{fec}
Inec: Historia del instituto nacional de estadística y censos (inec). Accessed
  on 11 July 2021 in
  URL:https://www.ecuadorencifras.gob.ec/historia-del-instituto-nacional-de-estadistica-y-censos-inec/

\bibitem{fpe}
Inei: Historia. Accessed on 11 July 2021 in
  URL:https://www.inei.gob.pe/historia/

\bibitem{opendataint}
International, O.K.: Global open data index. Accessed on 11 July 2021 in
  URL:http://2015.index.okfn.org/place/

\bibitem{whowebpage}
Organisation, T.W.H.: World health organisation (who) coronavirus disease
  (covid‐19) outbreak. Accessed on 11 July 2021 in URL:
  https://www.euro.who.int/en/health-topics/health-emergencies/coronavirus-covid-19/news/news/2020/3/who-announces-covid-19-outbreak-a-pandemic

\bibitem{whoplan}
Organisation, W.H.: Covid-19 strategy update. Accessed on 11 July 2021 in URL:
  https://www.who.int/docs/default-source/coronaviruse/covid-strategy-update-14april2020.pdf

\bibitem{Chire2020}
Saire, J.C., Panford-Quainoo, K.: Twitter interaction to analyze covid-19
  impact in ghana, africa from march to july (2020)

\bibitem{Dunn1981}
W, D.: Public policy. an introduction p.~14 (1981)

\end{thebibliography}

\end{document}